\documentclass[conference]{IEEEtran}
\usepackage{cite}
\usepackage[T1]{fontenc}
\usepackage{graphicx}
\usepackage{textcomp}
\usepackage{comment}
\usepackage{eso-pic} 

\def\BibTeX{{\rm B\kern-.05em{\sc i\kern-.025em b}\kern-.08em
    T\kern-.1667em\lower.7ex\hbox{E}\kern-.125emX}}

\begin{document}
\AddToShipoutPictureFG*{%
  \AtPageLowerLeft{%
    \put(\LenToUnit{0.5\paperwidth},\LenToUnit{0.32in}){%
      \makebox[0pt][c]{\parbox{\textwidth}{\centering\scriptsize
        \textcopyright~2026 IEEE. Personal use of this material is permitted. Permission from IEEE must be obtained for all other uses, in any current or future media, including reprinting/republishing this material for advertising or promotional purposes, creating new collective works, for resale or redistribution to servers or lists, or reuse of any copyrighted component of this work in other works.\\
        Accepted to the 2026 IEEE Frontiers in Education Conference (FIE). DOI: TBD}}}}}

\title{Understanding Student Use of Large Language Models Across Computer Science Subfields}

\author{
\IEEEauthorblockN{
Sehrish Basir Nizamani,
Yoonje Lee,
Nikitha Donekal Chandrashekar,
Margaret Ellis,
and Naren Ramakrishnan
}
\IEEEauthorblockA{
Department of Computer Science\\
Virginia Tech\\
Blacksburg, VA, USA\\
\{sehrishbasir, ylee201, nikitha, maellis1, naren\}@vt.edu
}
}

\maketitle

\begin{abstract}
This research full paper examines how undergraduate students use large language models (LLMs) across computer science subfields. As LLMs become increasingly integrated into computing education, understanding how their use varies across technical and pedagogical contexts is essential for designing effective, subfield-aware instruction. This paper presents a cross-subfield analysis of LLM usage among 211 undergraduate students in a problem-solving course intentionally designed to support responsible and effective LLM use through structured instruction and reflection.

Using post-assignment reflection data collected across seven instructional modules spanning multiple computer science subfields, we examine prompt counts, LLM role conceptualization, and verification behavior. Results show that LLM adoption varies substantially by assignment, with higher usage in algorithms and web development and lower usage in software engineering. Students predominantly treat LLMs as assistive tools rather than authoritative sources, and verification is common across all subfields, with most students using multiple strategies. Verification behavior also varies by assignment context, with testing more common in structured tasks and web search more common in open-ended tasks. 

These findings suggest that assignment characteristics play a central role in shaping how students interact with and evaluate LLM outputs, even under a single, consistently applied instructional design. This work contributes empirical evidence on how LLM adoption, role conceptualization, and verification behavior vary across computer science subfields, extending our prior work on structured, reflective LLM instruction to show how its effects differ by task rather than only in aggregate.
\end{abstract}



\section{Introduction}

Large language models (LLMs) have become deeply embedded in how students approach problem solving across computer science, and the pace of research responding to this shift has been striking. Within just the past few years, computing researchers have pursued markedly different strategies for understanding and shaping this adoption: some have focused on measuring what LLMs can do on isolated programming tasks \cite{chen2021codex, finnieansley2022robots}; others have studied LLM use within specific collaborative or design-oriented settings \cite{dhillon2024cowriting, zhou2026ux, kharrufa2026teamprojects}; and a growing body of work has turned toward AI literacy as a lens for equipping students with the conceptual grounding to use these tools well \cite{knoth2024ailiteracy}. Despite this breadth of activity, most of it treats computer science as a relatively uniform domain, rarely asking whether student LLM use looks different when the task shifts from, say, debugging an algorithm to designing a user interface.
In our own prior work, we developed and evaluated a \textit{demystify, use, and reflect} intervention within a required post-CS1 course spanning multiple subfields \cite{donekal2026demystify}.
That intervention opened up how LLMs work to students, guided their use through prompting strategies, and built in structured reflection; we found that students' self-reported understanding grew more technical while their verification habits became more deliberate and collaborative. However, that study characterized these shifts only at the level of the course as a whole. It left open a more granular question: given that the intervention spans subfields as different as algorithms, software engineering, and human-computer interaction, does its effect on student LLM use actually look the same across all of them, or does it vary with the nature of the task?

This paper takes up that gap directly. Building on the same intervention, we present a cross-subfield analysis of student LLM usage across the seven instructional modules analyzed in this study: algorithms, software engineering, networking, web development, human-computer interaction (HCI), databases, and machine learning. Specifically, we address the following research questions:

RQ1: How does LLM adoption and prompt-count intensity vary across CS subfields under a shared instructional intervention?

RQ2: How do students conceptualize the LLM's role (assistant, peer, expert) across different CS subfields?

RQ3: How do students' verification strategies for LLM outputs vary across CS subfields?

This work makes 
three main contributions: (1) a cross-subfield analysis of LLM adoption and usage patterns across seven representative assignments spanning multiple computer science subfields within a single undergraduate course; (2) a characterization of how students conceptualize LLMs and verify their outputs through manual reasoning, testing, and web search; and (3) an extension of our prior demystify, use, and reflect intervention that reveals how its effects vary at the subfield level rather than only in aggregate. 

\section{Related Work}
 
\subsection{LLMs in Programming / CS Education}
 
Research on LLMs in computing education has largely centered on programming tasks and code generation, with the bulk of this work concentrated in introductory programming settings. Foundational studies show that LLMs can generate functional code, solving a majority of benchmark programming problems comparable in difficulty to introductory or easy interview-style questions \cite{chen2021codex}
and in one study scoring within the top quartile of real CS1 students on invigilated exams \cite{finnieansley2022robots}. Deployed classroom studies reinforce this same introductory focus: CodeAid, an LLM-based coding assistant, was deployed in a 12-week introductory programming course of roughly 700 students, with 372 students actively engaging the tool \cite{KazemitabaarCodeAid}, and a separate study found that GitHub Copilot solved roughly half of a set of introductory programming exercises on its first attempt, with prompt engineering resolving most of the rest \cite{DennyConversing}. In educational settings, researchers have raised concerns that AI-assisted coding tools may lead to challenges in understanding, debugging, and evaluating generated code \cite{finnieansley2023exam, becker2023programming}, though direct empirical evidence of these specific student-side struggles remains limited. At the same time, the developers of Codex, OpenAI's code-generation model, themselves note risks such as incorrect or incomplete solutions and potential over-reliance on AI-generated outputs \cite{chen2021codex}. 

To the best of our knowledge, most prior research still treats LLMs as coding assistants for isolated, introductory programming tasks, with little attention to how student use might differ across CS subfields within a single course. A few recent studies, though, are starting to look beyond that narrow frame. 
A particularly relevant recent study surveyed and interviewed students in a 7-week, full-time undergraduate software engineering team project and proposed a three-dimensional design space for educational generative AI (GenAI) tools: the role GenAI plays (e.g., educator, peer, or assistant), the level of scaffolding provided (e.g., fading support as students gain competence), and the transparency of AI use for both instructors and teammates \cite{kharrufa2026teamprojects}. 
That work is important because it treats LLM as part of instructional design rather than only as a coding aid.  
 
\subsection{Student Use, Perceptions, and Interaction with LLMs}
 
Existing research shows that students tend to view LLMs as flexible academic helpers rather than replacements for human expertise, using them for brainstorming, writing support, and information seeking \cite{chan2023voices}. Studies of prompt design further reveal that non-expert users often rely on trial-and-error prompting rather than systematic strategies, making LLM interaction brittle and difficult to control \cite{zamfirescu2023johnny}.
 
Work on AI literacy suggests students' understanding of model capabilities is one factor, among others, in prompting effectiveness \cite{knoth2024ailiteracy}. Complementing this perspective, research on human–AI collaboration suggests that the effectiveness of combining humans and AI depends on the type of task and the relative strengths of each, rather than consistently outperforming either humans or AI alone \cite{vaccaro2024humanai}.
 
How students frame an LLM's role also appears tied to broader beliefs about its legitimacy as a professional tool. 
Vadaparty et al. \cite{vadaparty2026cs1llm} found that in an LLM-integrated CS1 course, 43.8\% of students believed professional programmers use tools like Copilot only rarely or sometimes, while the remaining 56.2\% believed professionals use such tools routinely or daily. The same study separately found that roughly a quarter of students expressed concerns about being over-reliant on Copilot; the authors speculate, without directly testing the connection, that skepticism about Copilot's legitimacy as a professional tool may be one contributing factor. Notably, these overreliance concerns arose even though the course's overall exam performance was comparable to historical, pre-LLM benchmarks.

Individual differences shape this picture further: Margulieux et al. \cite{margulieux2024selfregulation} found that introductory programming students with higher self-efficacy or lower fear of failure tended to use generative AI less often and later in their problem-solving process than their peers, even though how much a student used AI was not related to their self-regulation strategies.
 
While these studies provide important insights into how students conceptualize and interact with LLMs, to the best of our knowledge, we have not identified these perceptions and interaction patterns across different types of computing tasks within the same student population. Our work aims to contribute to this space by analyzing role conceptualization and interaction intensity across multiple subfields.
 
\subsection{Verification, Trust, and Responsible AI Use}

A growing body of work has examined the reliability limitations of LLMs. Bender et al. \cite{bender2021stochastic} argue that language models operate by manipulating linguistic form without grounding in meaning or communicative intent, creating risk that fluent, coherent-seeming output is mistaken for trustworthy or intentional communication. This concern is borne out empirically: LLMs have a well-documented tendency to produce fluent but factually incorrect or unsupported content, commonly termed hallucination \cite{ALANSARI2026100970}. These limitations underscore the importance of human verification and oversight when using LLM-generated outputs.
 
Work grounded in metacognition theory helps explain why this miscalibration occurs. Tankelevitch et al. \cite{tankelevitch2024metacognitive} frame evaluating an AI-generated output as a distinct metacognitive demand, requiring users to hold well-adjusted confidence in their own ability to judge an output's validity rather than simply trusting the response's fluency. Consistent with this framing, Singh et al. \cite{singh2025metacognitive} found that participants' self-reported metacognitive awareness did not reliably predict their actual trust in generated outputs, with output evaluation instead depending heavily on whether a user had enough domain knowledge to judge what they were looking at. In programming contexts specifically, López-Pernas et al. \cite{lopezpernas2025selfregulation} observed that student-AI interactions tended to follow a reactive loop of requesting a feature, copying in the generated code, testing it, and reporting errors back, with comparatively little of the reflective, self-evaluative behavior that genuine verification requires.
 
\subsection{LLMs Beyond Programming: Design, HCI, and Domain Workflows}
 
Beyond programming, LLMs are increasingly studied as tools for design, writing, and broader domain workflows. 
In user experience (UX) and HCI, recent work shows that practitioners use LLMs across a range of design activities, including collecting information for product and market research, processing and summarizing that information, offering design guidelines, simulating stakeholders, constructing user profiles, generating research instruments such as surveys and interview outlines, ideating design concepts, assisting with prototyping, and assisting with UX writing, while also encountering challenges refining prompts and verifying unreliable or fabricated outputs \cite{zhou2026ux}.
Similarly, studies of human-AI co-writing highlight the potential of LLMs to support iterative content creation, alongside concerns about authorship, control, and collaboration quality \cite{dhillon2024cowriting}.

Work on genAI in other specific subfields beyond programming is following a similar trajectory. Khan et al. \cite{khan2026mlcourse} examined genAI use in an upper-level machine learning course, finding that students most often turned to a course-deployed AI assistant for clarifying course concepts rather than for direct programming help, suggesting that LLM use in upper-level courses may shift toward conceptual support as learning objectives move beyond programming. In database systems education, Riazi and Rooshenas \cite{riazi2025dbfeedback} developed and piloted an LLM-driven feedback system for conceptual database design, in which students submit entity-relationship diagrams and receive targeted, LLM-generated feedback on specific relationships. While most students rated the generated feedback favorably, roughly a third of all AI-generated feedback in their pilot prompted a follow-up discussion thread because it was unclear, inaccurate, or left students with further questions, with accuracy notably weaker for certain constraint types such as total participation and specialization or superclass relationships. This underscores that even LLM tools purpose-built for a specific CS subfield can require ongoing human oversight, a dynamic our study examines directly through students' own reported verification behavior across subfields.
 
This expansion reflects a broader shift in which LLMs are used across diverse computing and knowledge work tasks rather than solely for code generation. However, existing research in these areas remains fragmented, often focusing on specific subfield or tasks in isolation. There is limited understanding of how LLM use compares across multiple computing subfields within a shared instructional setting. Our work addresses this gap by providing a cross-subfield analysis of LLM interaction, role conceptualization, and verification strategies within a single course context.

\section{Course Context}
This study was conducted in an undergraduate problem-solving course in computer science at a large research-intensive (R1) university in the United States during Fall 2025. The course is intentionally designed to expose students to multiple computer science subfields while simultaneously supporting informed and responsible use of LLMs.

Rather than treating LLM use as incidental or unregulated, the course adopted a structured intervention centered on demystifying LLMs, how they work, their capabilities, and their probabilistic nature and limitations, alongside guided use through prompting strategies and iterative interaction. Rather than explicitly encouraging LLM use, the course left the decision of whether and how to use them up to each student. All students had access to Microsoft Copilot Chat through an institutional license available at the time of the study, though students were not restricted to a single tool and were free to use any LLM of their choosing. The course also emphasized ethical use, requiring students to understand, explain, and acknowledge any AI-generated contributions. This design positions LLMs as tools for supported learning rather than replacements for student reasoning.

Table~\ref{tab:course} summarizes the course structure across subfields, including core topics and the role of LLMs in each module.

\begin{table*}[htbp]
\caption{Course Structure Across Subfields}
\label{tab:course}
\centering
\begin{tabular}{|p{2.8cm}|c|p{3.8cm}|p{5.2cm}|}
\hline
\textbf{CS Subfields} & \textbf{Weeks} & \textbf{Core Topics} & \textbf{LLM Intervention} \\
\hline
Algorithms (ALG) & 1--4 & Python basics, prompting, algorithms, recursion, regular expressions, testing & LLM-assisted coding, debugging, and test-based verification \\
\hline
Software Engineering (SWE) & 5--6 & Software process, design, version control (Git) & Evaluating LLM reliability in structured development workflows \\
\hline
Systems / Networking (NETWORK) & 7--8 & Internet architecture, data centers, HTTP protocols & LLM use for conceptual understanding and implementation support \\
\hline
Web Development (WEB) & 9 & HTML, containers, web services & LLM-assisted implementation and debugging of web components \\
\hline
HCI / UX (HCI) & 10--11 & Usability, design, prototyping & LLM-supported ideation, interpretation, and design reasoning \\
\hline
Databases (DB) & 12 & Databases, SQL & LLM use for querying and data reasoning \\
\hline
Data Science / Machine Learning (ML) & 13 & Data science, machine learning concepts and models & LLM use for conceptual understanding and prediction tasks \\
\hline
\end{tabular}
\end{table*}

For this study, one representative assignment was selected from each subfield. These assignments span a range of task types, including programming, system interaction, and open-ended design. In algorithms, students implemented and tested recursive and dynamic programming solutions to the knapsack problem. In software engineering, students practiced collaborative version control using Git. Networking tasks involved constructing and testing HTTP (HyperText Transfer Protocol) requests, while web development assignments focused on building and styling interactive HTML (HyperText Markup Language) components. In HCI, students produced a usability engineering report including personas, prototypes, and heuristic evaluations. Database assignments required querying and processing structured data using SQL (Structured Query Language) and Python, and machine learning tasks involved training and applying a Naive Bayes classifier.

All assignments were completed by the same group of students under consistent instructional conditions, enabling within-course comparison of LLM usage across computer science subfields while minimizing variability due to differences in student populations or learning environments. Instruction throughout the semester emphasized effective prompting strategies, iterative interaction, and assessing the reliability of generated responses.

A key component of this design was structured reflection. After each assignment, students completed a short reflection survey describing their LLM use (Fig.~\ref{fig:survey}). By capturing interaction immediately after task completion, these reflections enable systematic analysis of how students engage with LLMs across diverse computing contexts under consistent instructional scaffolding, rather than ad hoc or unregulated use.

Together, this design provides a controlled setting in which the same students engage with LLMs across multiple subfields using consistent reflective instruments, supporting robust within-subject analysis of interaction patterns.

\section{Methodology}

\subsection{Participants}

Participants were undergraduate students enrolled in the course (N = 367). Of these, 90.5\% (n = 332) provided informed consent to participate in the study. Students were allowed to drop up to two classwork assignments and one homework assignment during the semester without penalty, and we included only those who completed all seven assignments analyzed, yielding a final sample of 211 students (63.6\% of consenting participants). Some of the remaining attrition may reflect students completing an assignment but not the accompanying reflection survey.

\subsection{Study Design}

Each representative assignment serves as the unit of analysis, corresponding to one instructional module. The study follows a within-subject design, as the same students completed all assignments under consistent instructional conditions. This enables direct comparison of LLM usage across diverse computing contexts while minimizing variability due to differences in student populations or learning environments. Table~\ref{tab:assignments} lists the representative assignment and task description selected for each subfield.

\begin{table*}[t]
\caption{Representative Assignments and Task Descriptions Across Subfields}
\label{tab:assignments}
\centering
\begin{tabular}{|p{3cm}|p{5cm}|p{8cm}|}
\hline
\textbf{CS Subfield} & \textbf{Assignment Used in Analysis} & \textbf{Task Description} \\
\hline
Algorithms (ALG) & Homework: Knapsack Problem Comprehension & Students implemented recursive and dynamic programming solutions, wrote unit tests, and analyzed constraints such as conflicting items and bonus combinations. \\
\hline
Software Engineering (SWE) & Classwork: Git Commands Practice Day & Students practiced version control by creating repositories, committing changes, and collaborating with peers using Git. \\
\hline
Networking (NETWORK) & Classwork: HTTP Request Methods & Students sent and retrieved data using POST and GET requests through Postman, curl, and Python. \\
\hline
Web Development (WEB) & Classwork: HTML Table Design & Students created and styled an interactive HTML table using CSS, including hover effects and sticky headers. \\
\hline
HCI / UX (HCI) & Homework: Usability Engineering Report & Students documented personas, wireframes, prototypes, and heuristic evaluations, emphasizing iterative usability improvements. \\
\hline
Databases (DB) & Classwork: SQL and Python & Students queried an SQLite database using Python, applied filtering and sorting, and generated HTML output. \\
\hline
Machine Learning (ML) & Classwork: ML Training and Prediction & Students trained a Naive Bayes model on labeled email data and classified new messages as spam or ham. \\
\hline
\end{tabular}
\end{table*}

\subsection{Data Source and Measures}

After completing each assignment selected for this study, students completed a short reflection survey describing their use of LLMs for that assignment.

\begin{figure}[htbp]
\centering
\includegraphics[width=\columnwidth]{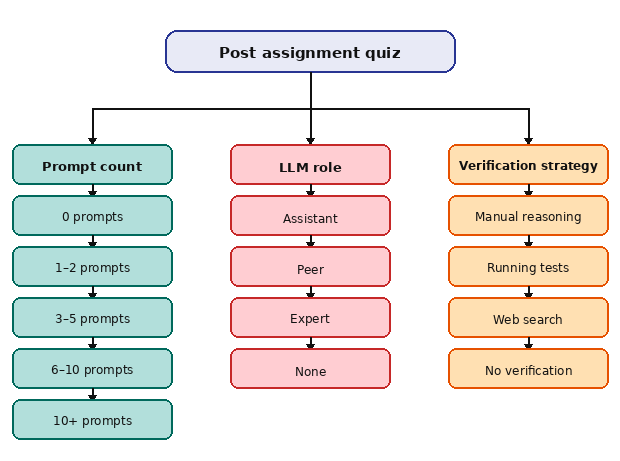}
\caption{Structure of post-assignment reflection survey capturing LLM interaction, role conceptualization, and verification strategies}
\label{fig:survey}
\end{figure}

The reflection survey captured three aspects of LLM interaction: prompt counts, conceptualization of the LLM’s role, and verification strategies. As shown in Fig.~\ref{fig:survey}, prompt count was measured using prompt count ranges (0, 1--2, 3--5, 6--10, 10+). Students also reported how they conceptualized the LLM (assistant, peer, expert, or none) and selected one or more strategies used to verify LLM-generated outputs, including manual reasoning, testing, web search, or no verification. Both the role categories and the verification strategies build on patterns of LLM engagement identified in our team's earlier work \cite{donekal2026demystify}: thematic analysis of student reflections there found that students described LLM use ranging from a starting point for their own reasoning (an Assistant role) to collaborative engagement (a Peer role) to relying on it with minimal personal input (an Expert role), and separately noted a shift toward fact-checking and cross-referencing strategies alongside a decline in unverified use, patterns that informed both our role categories and our verification-strategy categories in the present study.

\subsection{Data Processing and Analysis}

Survey responses were exported from Canvas as separate CSV files for each assignment. Relevant columns were retained and standardized across files. Prompt count responses required normalization due to formatting inconsistencies in exported data (e.g., date-like values corresponding to prompt ranges).

The datasets were merged into a single structured dataset using student identifiers, where each row represents one student and columns correspond to responses across all assignments. This enabled consistent comparison of LLM usage across subfields.

We conducted descriptive analyses to examine patterns of LLM usage across assignments. Interaction turns were treated as ordinal categories (0, 1--2, 3--5, 6--10, 10+), LLM role as categorical (assistant, peer, expert, none), and verification methods as multi-select categorical variables.

To examine differences in interaction turns across subfields, we used a Kruskal--Wallis test. Verification strategies were analyzed using descriptive statistics due to their multi-select nature, including examination of individual strategies and common combinations.

\section{Results}

\subsection{LLM Adoption and Prompt Count Distribution (RQ1)}

Students reported how many prompts they wrote when using LLMs for each assignment. As shown in Table~\ref{tab:llm_adoption}, LLM adoption varied substantially by assignment, with Algorithms (89.6\%) and Web Development (80.6\%) showing the highest usage, and Software Engineering the lowest (15.2\%). Networking and Databases exhibited moderate adoption (58.3\%), while HCI (43.1\%) and Machine Learning (40.3\%) showed lower usage.


\begin{table}[htbp]
\caption{LLM Adoption Across Computer Science Subfields}
\label{tab:llm_adoption}
\centering
\begin{tabular}{lcc}
\hline
\textbf{Subfield} & \textbf{Used LLM (\%)} & \textbf{No Use (\%)} \\
\hline
Algorithms & 89.6 & 10.4 \\
Web Development & 80.6 & 19.4 \\
Networking & 58.3 & 41.7 \\
Databases & 58.3 & 41.7 \\
HCI & 43.1 & 56.9 \\
Machine Learning & 40.3 & 59.7 \\
Software Eng. & 15.2 & 84.8 \\
\hline
\end{tabular}
\end{table}

Among students who used LLMs, moderate prompting was the most common pattern. Across assignments, 3--5 prompts accounted for approximately 29-46\% of users, while 6-10 prompts accounted for 13-34\%. Extended prompting of 10+ prompts was rare except in Algorithms (18.0\%). Minimal use of 1-2 prompts was more common in Machine Learning and Software Engineering than in other assignments. A Kruskal–Wallis test confirmed that prompt count distributions differed significantly across assignments (H = 71.18, df = 6, p $<$ 0.001).

\subsection{LLM Role Conceptualization (RQ2)}
Students categorized their LLM use as Assistant, Expert, Peer, or None. 
Table~\ref{tab:llm_usage_roles} summarizes the distribution of LLM usage roles across subfields, computed over all students ($N = 211$ per assignment).

\begin{table}[htbp]
\caption{LLM Usage Roles Across Computer Science Subfields (All Students)}
\label{tab:llm_usage_roles}
\resizebox{\columnwidth}{!}{%
\begin{tabular}{lccccc}
\hline
\textbf{Subfield} & \textbf{Assistant} & \textbf{Expert} & \textbf{Peer} & \textbf{N/A} & \textbf{N} \\
 & \textbf{n (\%)} & \textbf{n (\%)} & \textbf{n (\%)} & \textbf{n (\%)} & \\
\hline
Algorithms & 130 (61.6) & 19 (9.0) & 38 (18.0) & 24 (11.4) & 211 \\
Databases & 85 (40.3) & 13 (6.2) & 25 (11.8) & 88 (41.7) & 211 \\
HCI & 67 (31.8) & 1 (0.5) & 22 (10.4) & 121 (57.3) & 211 \\
Machine Learning & 60 (28.4) & 7 (3.3) & 18 (8.5) & 126 (59.7) & 211 \\
Networking & 94 (44.5) & 7 (3.3) & 20 (9.5) & 90 (42.7) & 211 \\
Software Eng. & 26 (12.3) & 0 (0.0) & 3 (1.4) & 182 (86.3) & 211 \\
Web Development & 133 (63.0) & 19 (9.0) & 18 (8.5) & 41 (19.4) & 211 \\
\hline
\end{tabular}%
}
\end{table}

In Table~\ref{tab:llm_usage_roles}, 'N/A' includes students who reported zero prompts as well as a small number of students per assignment (1--3) who reported a nonzero prompt count but selected 'None' as their LLM role.

Across all assignments, Assistant usage was the most common LLM role among students who used LLMs. This pattern is most visible in Algorithms (61.6\%) and Web Development (63.0\%), where a majority of students reported using LLMs in an assistive role.

Peer usage was consistently lower than Assistant usage but remained the second most common role in most assignments. It was most prominent in Algorithms (18.0\%) and Databases (11.8\%), with lower proportions in other subfields.

Expert usage was relatively uncommon across all assignments, ranging from 0.0\% in Software Engineering to 9.0\% in Algorithms and Web Development.

A substantial proportion of students reported no LLM use in several assignments, particularly in Software Engineering (86.3\%), HCI (57.3\%), and Machine Learning (59.7\%). 
In contrast, non-use was lowest in Algorithms (11.4\%) and Web Development (19.4\%), reflecting higher overall adoption in these subfields.

 Overall, the distribution of roles (assistant, peer, expert, none)  reflects both differences in LLM adoption and variation in how students engaged with LLMs across assignments. A chi-square test of independence confirmed that this distribution differed significantly across assignments ($\chi^2$ = 358.73, df = 18, p < 0.001).

\subsection{Verification Behavior (RQ3)}

This analysis includes only students who reported using LLMs for these assignments by selecting either assistant, peer, or expert for each assignment, ensuring that verification behavior reflects active LLM use. Students could select multiple verification methods, including manual reasoning, testing, web search, or no verification. Because multiple methods were allowed, percentages may sum to more than 100\%.

\begin{figure*}[t]
\centering
\includegraphics[width=\textwidth]{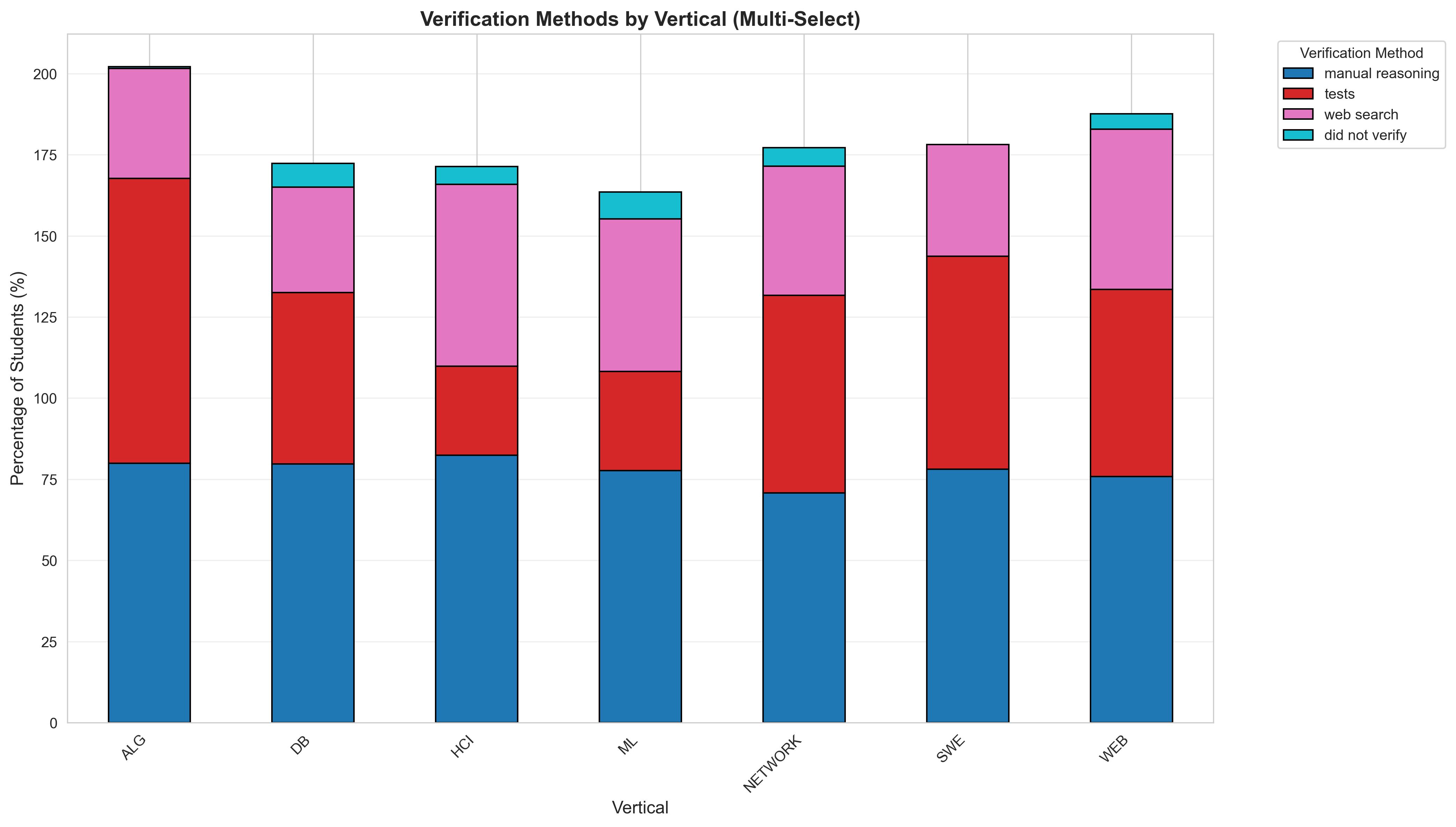}
\caption{Verification methods across subfields (LLM users only).}
\label{fig:verification_methods}
\end{figure*}

As shown in Fig.~\ref{fig:verification_methods}, manual reasoning was the most consistently used verification method, ranging from 70.7\% to 82.4\% across assignments. HCI showed the highest manual reasoning rate (82.4\%), while Networking had the lowest (70.7\%). 

This consistency suggests that students relied on their own reasoning regardless of assignment type, a pattern confirmed by a chi-square test showing no significant difference across assignments ($\chi^2$ = 5.73, df = 6, p = 0.455).

Testing behavior varied across assignments. Algorithms had the highest testing rate (87.8\%), while HCI had the lowest (27.5\%). Networking (61.0\%), Software Engineering (65.6\%), and Web Development (57.6\%) showed moderate testing usage, whereas Machine Learning (30.6\%) and HCI (27.5\%) showed lower reliance on testing. 
This variation was statistically significant ($\chi^2$ = 133.03, df = 6, p < 0.001).

Web search usage also varied across assignments, ranging from 32.5\% in Databases to 56.0\% in HCI. Higher web search usage was observed in HCI (56.0\%), Web Development (49.4\%), and Machine Learning (47.1\%), indicating greater reliance on external information sources in more conceptual or open-ended tasks. 
This variation was also statistically significant ($\chi^2$ = 22.79, df = 6, p < 0.001).

Very few students reported that they did not verify LLM outputs. Non-verification ranged from 0\% to 8\%, with Software Engineering showing no instances of unverified usage. Across all assignments, the LLM users (93\%--100\%) reported using at least one verification method, indicating that students did verify LLM outputs in most cases. Non-verification rates varied significantly across assignments ($\chi^2$ = 13.95, df = 6, p = 0.030), though all rates remained low.

\begin{figure*}[htbp]
\centering
\includegraphics[width=0.7\textwidth]{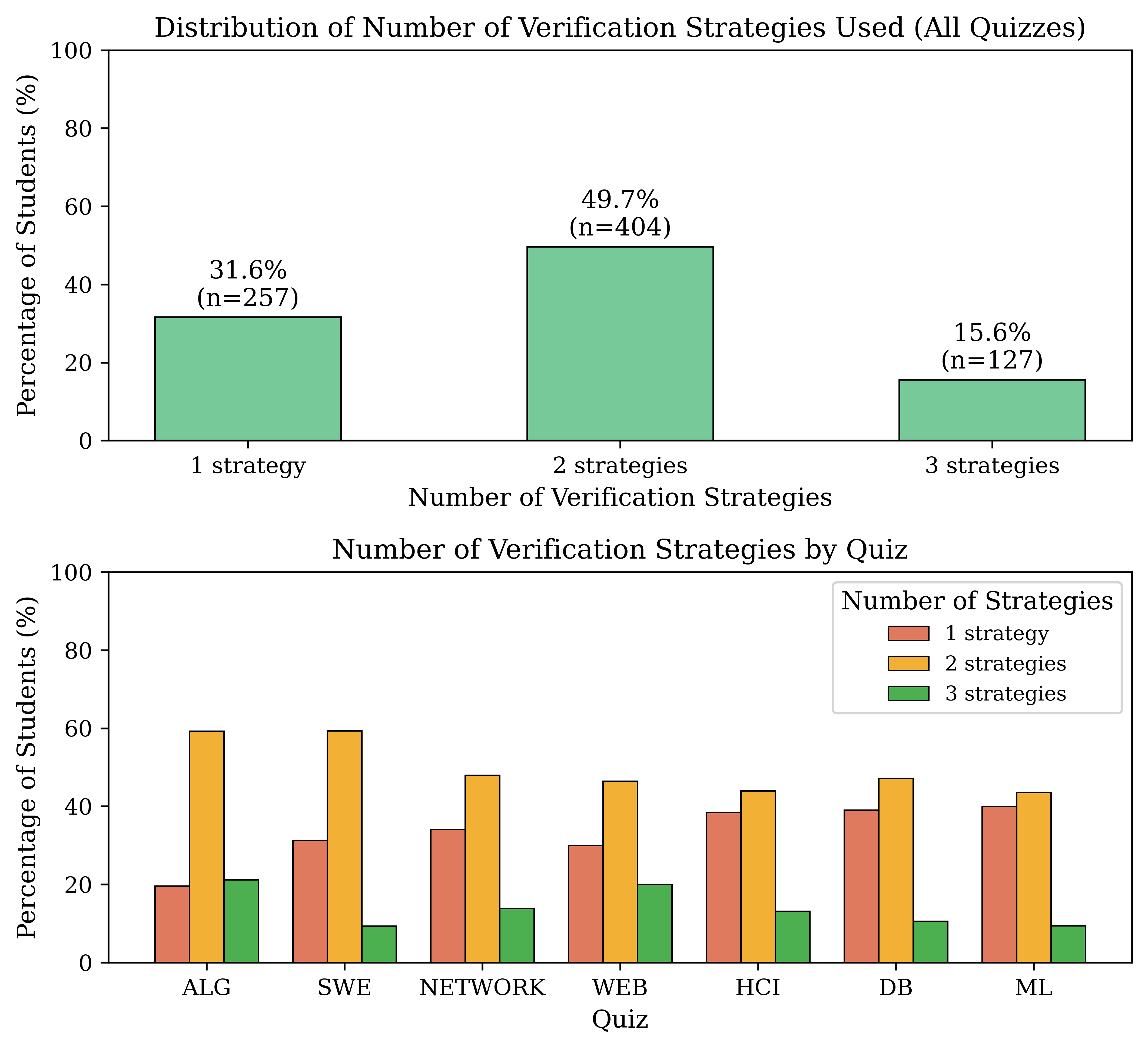}
\caption{Distribution of number of verification strategies used (LLM users only).}
\label{fig:strategy_count}
\end{figure*}

Fig.~\ref{fig:strategy_count} shows the number of verification strategies used by students. 
Most students used multiple strategies: 49.7\% used exactly two methods, 31.6\% used a single method, and 15.6\% used three methods (a further 3.1\% reported no verification method). This indicates that students typically combined verification approaches rather than relying on a single strategy.

Further analysis of verification strategy combinations showed that the most common pattern was ``manual reasoning + tests'' (27.5\%), followed by ``manual reasoning'' alone (18.8\%), and ``manual reasoning + tests + web search'' (15.6\%). Combinations involving web search were also common, particularly in assignments such as HCI and Machine Learning.

\section{Discussion}
Across the seven tasks, LLM use was highest for assignments we characterize as technically demanding, time-consuming, or requiring rapid iteration, and lowest for assignments that were procedural, highly structured, or already supported by detailed instructions. We interpret this pattern as reflecting differences in assignment complexity and instructional scaffolding; since we did not measure these factors directly, we cannot rule out other explanations for the observed differences.

The strongest example is Algorithms, where the Knapsack Problem Comprehension homework required recursive and dynamic programming solutions, unit testing, and reasoning about additional constraints such as conflicting items and bonus combinations. This kind of task is difficult, testable, and iterative, which likely made LLMs attractive as a coding and debugging aid. The high adoption rate and strong reliance on testing in this assignment support our interpretation. 

Web Development showed a similar pattern. Although the HTML Table Design task was not algorithmically complex, it required students to satisfy multiple CSS (Cascading Style Sheets) and layout constraints, manage hover and sticky-header behavior, and ensure that the final output matched the specification. The combination of visual detail, implementation accuracy, and debugging needs likely contributed to the high LLM adoption observed in this assignment.

In contrast, Software Engineering showed the lowest adoption. The Git Commands Practice Day assignment was comparatively procedural and strongly scaffolded, with explicit instructions for creating repositories, committing changes, and collaborating through version control. One plausible explanation is that clearly specified tasks reduce students' perceived need for AI assistance; however, task clarity was not measured directly and this pattern is drawn from a single assignment, so other unmeasured features may account for it.

The remaining assignments fall between these extremes. Networking and Databases showed moderate adoption and mixed verification behavior. These tasks were structured enough to support testing and checking, but they still required students to understand APIs, SQL queries, or data flow, which may have encouraged selective LLM use for clarification or implementation support, echoing the mixed reliability found in purpose-built LLM feedback tools for database design \cite{riazi2025dbfeedback}. HCI and Machine Learning, by contrast, were more open-ended and less directly testable. In these assignments, students were less likely to use LLMs as heavily, but when they did, they relied more on manual reasoning and web search. This pattern suggests that students may have found LLMs useful for idea generation or explanation, but not sufficient on their own for resolving ambiguous design or conceptual questions, consistent with prior findings in both HCI \cite{zhou2026ux} and machine learning \cite{khan2026mlcourse} contexts specifically.

The dominance of the Assistant role over Expert use among LLM users suggests that students were not simply delegating problem solving to the model. Instead, they appear to have used LLMs selectively to support their own work, especially in assignments that were difficult, iterative, or implementation-heavy. The prevalence of verification, particularly through manual reasoning and testing, reinforces this interpretation. Students' self-reported verification behavior suggests they were not simply accepting generated answers without checking them, and that this checking varied with the demands of the assignment.

The verification results also support the idea that assignment structure influenced how students judged LLM reliability. In structured and executable tasks, especially Algorithms and Web Development, testing was a natural and readily available verification method. In more open-ended tasks such as HCI and Machine Learning, web search became more prominent, likely because students needed external references to evaluate design choices or conceptual claims that could not be checked through tests alone. This difference suggests that the form of verification is closely tied to the type of problem being solved.

Taken together, we interpret these differences as an interaction between task demands and student decision-making, shaped by assignment complexity, verifiability, and scaffolding, rather than as a property of the subfield alone.

\section{Limitations}

This study is based on self-reported reflections, which may be subject to recall bias or reporting inaccuracies. This risk is greatest for students with heavier LLM use, who must distinguish between closely spaced ordinal prompt-count ranges (e.g., 3--5 versus 6--10) well after completing the assignment. Future work could validate self-reported counts against server-side usage logs where available.

A related measurement limitation concerns the LLM role categories (Assistant, Peer, Expert, None). Students selected among these labels without an accompanying in-survey definition, and the instrument was not formally piloted to confirm consistent interpretation. As a result, we cannot verify whether a student's self-reported role reflects a genuine difference in how the LLM was used (e.g., a starting point for reasoning versus a complete solution) or simply a differing interpretation of the label itself. Future iterations should provide explicit role definitions and validate self-reported categorizations against independent evidence, such as prompt logs.

Approximately 36\% of consenting participants were excluded for not completing all seven reflection surveys. While the course's standard drop policy (up to two classwork and one homework assignment) accounts for some of this attrition, we cannot rule out that assignment difficulty or attitudes toward LLM use also contributed to non-completion, a factor we did not measure directly.

The findings also reflect a single course context with a specific set of assignments and instructional practices; results are tied to the assignments used here and should not be generalized without caution to all tasks in the same subfields.

Finally, the study identifies associations between assignment characteristics and LLM usage patterns but does not establish causality. Future work should examine how specific design features, such as assignment clarity, testability, and open-endedness, independently shape student interaction with LLMs.

\section{Conclusion}

This paper examined how undergraduate students use LLMs across multiple computer science subfields within a structured problem-solving course, addressing three research questions concerning LLM adoption and prompt-count intensity (RQ1), role conceptualization (RQ2), and verification behavior (RQ3) across seven assignments and 211 students.

Results show that LLM use varies significantly by assignment type, with higher adoption in algorithmic and implementation-oriented tasks and lower usage in software engineering, where the assignment used in this study, Git Commands Practice Day, was comparatively procedural and strongly scaffolded (see Discussion). Students predominantly treated LLMs as assistive tools rather than authoritative sources, and verification was nearly universal, with most students employing multiple strategies. Verification approaches also differed by subfield, with testing more common in structured tasks and web search more common in open-ended contexts.

These findings highlight the importance of assignment design in shaping how students engage with LLMs. Rather than restricting LLM use, educators can design tasks and reflection activities that promote effective and critical interaction with AI tools. This work contributes to understanding how LLMs can be integrated into computing education in a way that supports responsible and meaningful learning.

\section*{Acknowledgment}
This work was supported in part by the Virginia Tech Center for Excellence in Teaching and Learning (CETL) through a High-Impact Project Grant, which we gratefully acknowledge for supporting the development of the course and instructional activities that made this research possible. LLMs were used to assist with writing Python scripts for data analysis and with structuring portions of this manuscript. All analyses, interpretations, and conclusions were developed by the authors, who manually reviewed, verified, and edited all AI-assisted code and content, and take full responsibility for the accuracy and integrity of this work.

\bibliographystyle{IEEEtran}
\bibliography{references}

\end{document}